\documentclass[conference,twocolumn]{IEEEtran}
\usepackage{amsfonts}
\usepackage{times}
\usepackage{graphicx}
\usepackage{latexsym}
\usepackage{dsfont}
\usepackage{amssymb}
\usepackage{amsmath}
\usepackage{cite}
\usepackage{verbatim}
\usepackage{subfigure}

\newcommand{\figref}[1]{{Fig.}~\ref{#1}}

\def\bb0{{\mathbb{0}}}

\def\ba{{\mathbf{a}}}
\def\bb{{\mathbf{b}}}

\def\bg{{\mathbf{g}}}
\def\bh{{\mathbf{h}}}

\def\bm{{\mathbf{m}}}

\def\bw{{\mathbf{w}}}

\def\by{{\mathbf{y}}}
\def\bz{{\mathbf{z}}}
\def\b0{{\mathbf{0}}}

\def\bA{{\mathbf{A}}}

\def\bR{{\mathbf{R}}}

\def\bU{{\mathbf{U}}}

\def\bbE{{\mathbb{E}}}

\def\cA{\mathcal{A}}

\def\cN{\mathcal{N}}

\def\sf0{{\mathsf{0}}}

\usepackage{epstopdf}
\usepackage{enumerate}
\usepackage{algorithmicx}
\usepackage{algorithm}
\usepackage{amsmath}
\usepackage[noend]{algpseudocode}
\usepackage{float}
\usepackage{hyperref}
\usepackage{color}
\usepackage{makeidx}
\usepackage{bbm}
\usepackage{graphicx}
\usepackage{etoolbox}

\AtBeginEnvironment{figure}{\setcounter{subcaption}{0}}%
\AtBeginEnvironment{table}{\setcounter{subcaption}{0}}%

\makeatletter
\newcounter{subcaption}

\makeatother
\newcommand{\sref}[1]{{Section}~\ref{#1}}

\begin{document}
\title{Generative Adversarial Estimation of Channel Covariance in Vehicular Millimeter Wave Systems}
\author{ Xiaofeng Li, Ahmed Alkhateeb, and Cihan Tepedelenlio\u{g}lu\\ Arizona State University, Email: xiaofen2, aalkhateeb, cihan@asu.edu}
\maketitle

\begin{abstract}
Enabling highly-mobile millimeter wave (mmWave) systems is challenging because of the huge training overhead associated with acquiring the channel knowledge or designing the narrow beams. Current mmWave beam training and channel estimation techniques do not normally make use of the prior beam training or channel estimation observations.
Intuitively, though, the channel matrices are functions of the various elements of the environment. Learning these functions can dramatically reduce the training overhead needed to obtain the channel knowledge.
In this paper, a novel solution that exploits machine learning tools, namely conditional generative adversarial networks (GAN), is developed to learn these functions between the environment and the channel covariance matrices. More specifically, the proposed machine learning model treats the covariance matrices as 2D images and learns the mapping function relating the uplink received pilots, which act as RF signatures of the environment, and these images. Simulation results show that the developed strategy efficiently predicts the covariance matrices of the large-dimensional mmWave channels with negligible training overhead.

\end{abstract}

\section{Introduction} \label{sec:Intro}
Millimeter wave (mmWave) communication is a promising technology for the applications that demand high data rates with high mobility such as vehicular communications and wireless virtual/augmented reality \cite{Alkhateeb2018}. Enabling these highly-mobile applications, though, requires developing efficient techniques that acquire the large mmWave channels with low training overhead. Since estimating the full channel matrices every coherence time may not be feasible in these highly-mobile systems, a reasonable approach is to obtain long-term channel statistics such as spatial channel covariance, which can then be leveraged for both the channel estimation and the precoder design \cite{Adhikary2014,park2017spatial}. 

Prior work on mmWave channel covariance estimation leveraged the sparse nature of the channels and developed compressive sensing based techniques  \cite{lee2016channel,park2017spatial}. While these techniques can generally reduce the training overhead compared to exhaustive search solutions, this overhead is still large for large array systems and scales with the number of antennas. Further, compressive estimation techniques normally make hard assumptions on the exact sparsity of the channels which render their practical feasibility questionable. 

In this paper, we propose a novel solution for highly-mobile mmWave systems that leverages deep learning tools to efficiently learn and predict the mmWave channel covariance matrices. The key idea of this solution is to treat the covariance matrices as 2D images where GAN networks can be trained to learn the important features of these images. More specifically, following \cite{Alkhateeb2018}, the developed solution requires the mobile user to transmit only one uplink training sequence that gets received jointly by multiple base stations (BSs) using \textit{omni} beam patterns, i.e., with negligible training overhead. These received training signals represent an RF signature of both the environment and the transmitter/receiver locations. A conditional GAN network is then leveraged to learn the implicit mapping function between the received training signals and a sparse representation of the channel covariance matrix. Simulation results, based on accurate ray tracing, show that the proposed solution can efficiently predict large-dimensional mmWave channel covariance matrices with small mean-squared errors. 

\textbf{Notation}: We use the following notation throughout this paper: $\bA$ is a matrix, $\ba$ is a vector, $a$ is a scalar, and $\cA$ is a set. $|\bA|$ is the determinant of $\bA$, whereas $\bA^T$, $\bA^H$, are its transpose and Hermitian (conjugate transpose).  $\cN(\bm,\bR)$ is a complex Gaussian random vector with mean $\bm$ and covariance $\bR$. $\bbE\left[\cdot\right]$ is used to denote expectation.

\section{System and Channel Models} \label{sec:Model}
\begin{figure}[t]
	\centerline{
		\includegraphics[width=.9\columnwidth]{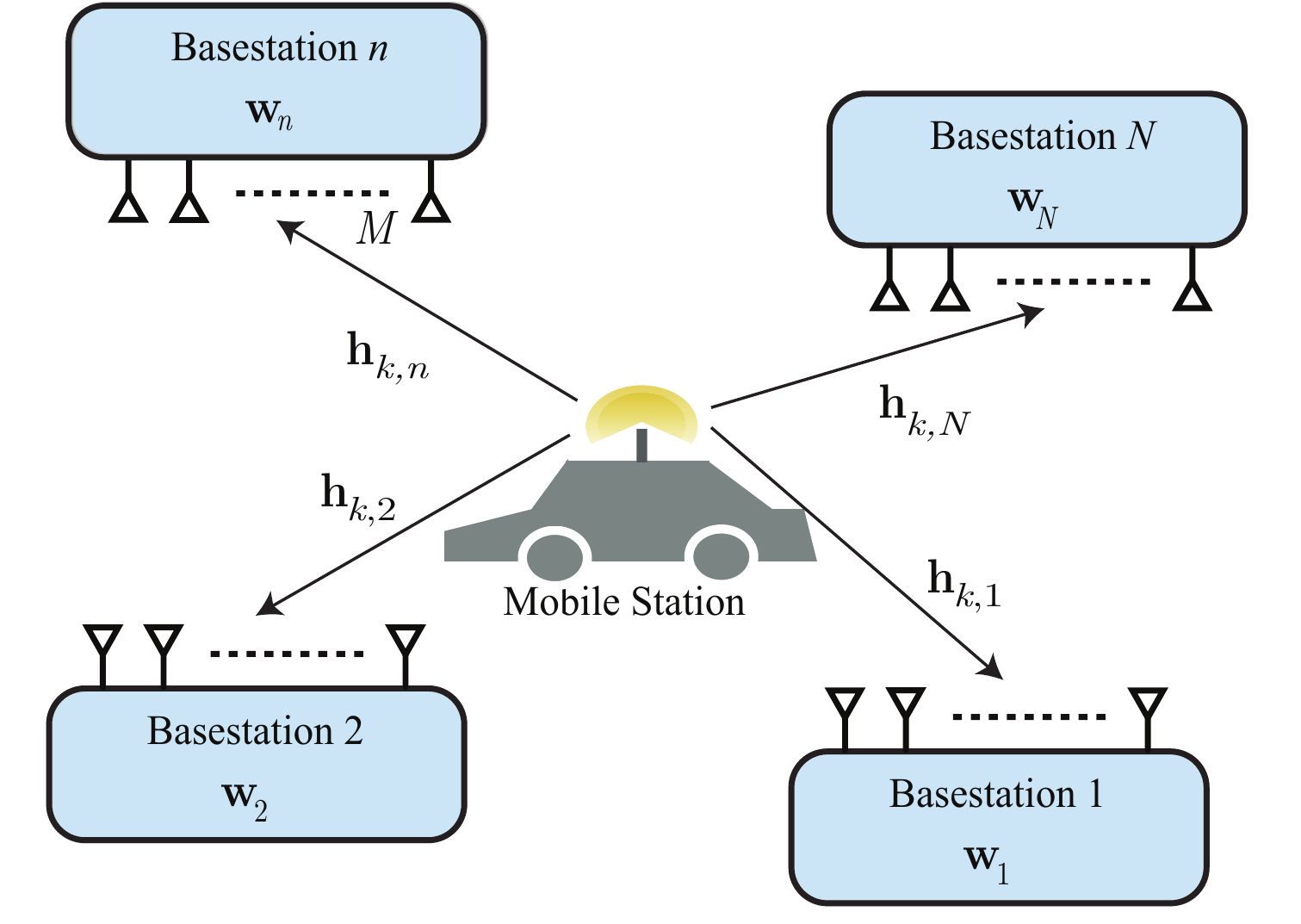}
	}
	\caption{A block diagram of the proposed mmWave system where $N$ BSs are installed in a street and simultaneously receive uplink training signals from one vehicular/mobile user. Each BS is equipped with $M$ antennas and one RF chain, and is applying analog-only beamforming/combining. The BSs are assumed to be connected together to share the received uplink training signals.}
	\label{fig:Sys_Model}
\end{figure}

In this section, we describe the adopted mmWave system and channel models. 

\subsection{System Model} \label{sec:Sys_Model}
Consider the mmWave communication system shown in Fig. \ref{fig:Sys_Model}, where $N$ base stations (BSs) are simultaneously serving one mobile user. Each BS is equipped with $M$ antennas which form a uniform linear array and the user has only one antenna. The BSs are assumed to be connected to each other so that they can share the uplink training signals, received from the mobile user. For simplicity, we assume that each BS has only one RF chain and is applying analog-only combining via a network of phase shifters during the uplink transmission \cite{HeathJr2016}.

Considering a wideband OFDM system, the uplink training symbol $s_k\in\mathbb{C}$ at subcarrier $k, k=1,\cdots,K$ is transformed to the time domain by a $K$-point IFFT. A cyclic prefix is then added to the symbol block to generate the transmit signal. Let $\tilde{\bh}_{k,n}\in\mathbb{C}^{M\times 1}$ denote the uplink channel vector between the user and the $n$th BS at the $k$th subcarrier, the post-combining received signal at subcarrier $k$ at BS $n$ can then be expressed as
\begin{align}
y_{k,n}=\bw_n^T\bh_{k,n} s_k + v_k,
\end{align}
where $\bw_n$ is the analog combiner at BS $n$, and $v_k \sim \cN(0,\sigma^2)$ is the Gaussian noise corrupting the received signal.

\subsection{Channel Model}
A wideband geometric channel model with $L$ clusters is adopted for our mmWave system \cite{alkhateeb2016frequency}. In this model, each of the clusters contributes with one ray which has a time delay $\tau_{n,\ell}$, and an angle of arrival (AoA) $\theta_{n,\ell}$. If  $p(t)$ denotes the pulse shaping function, the delay-d channel vector between the user and $n$th BS can be written as 
\begin{align}\label{eqn:channel_time}
\mathsf{\boldsymbol{h}}_{d,n}=\sqrt{\frac{M}{\rho}}\sum_{\ell=1}^{L} g_{n,\ell} p(dT_s-\tau_{n,\ell})\ba(\theta_{n,\ell})
\end{align}
where $\ba(\theta_\ell)$ is the array response vector at the AoA $\theta_{n,\ell}$, $\rho$ denotes the path-loss between BS $n$ and the user, and $g_{n,\ell}$ is the complex gain for the $\ell$th path. 
Given the time domain channel in (\ref{eqn:channel_time}), the frequency domain channel vector at subcarrier $k$ can be expressed as follows:
\begin{align}
{\bh}_{k,n}=\sum_{d=0}^{D-1} \mathsf{\boldsymbol{h}}_{d,n}e^{-j\frac{2\pi k}{K}d}.
\end{align}

\noindent Finally, the spatial channel covariance matrix is expressed as $\bR_{\bh_n}=\bbE[{\bh}_{k,n} {\bh}_{k,n}^H]$.

\section{Problem Definition} \label{sec:Problem}
The main objective of this paper is to develop an efficient solution for the mmWave channel covariance estimation $\bR_{\bh_n}$ that requires very low training overhead. More specifically, considering the system model in \sref{sec:Sys_Model}, our problem is to efficiently estimate the channel covariance $\bR_{\bh_n}$ at the $n$th BS given the concatenated training sequence defined as
\begin{equation} \label{eq:rec_signals}
\by=[y_{1,1},\cdots,y_{K,1},y_{1,2},\cdots,y_{K,2},\cdots,y_{K,N}]^T,
\end{equation}
which is collected from all the coordinating BSs. 

The major challenge of estimating the channel covariance in mmWave systems is the large channels dimensions due to the large number of antennas. This results in very high training overhead which becomes a limiting factor for the operation of highly-mobile mmWave systems such as vehicular communications and wireless virtual/augmented reality \cite{Alkhateeb2018}. Prior work  \cite{lee2016channel,park2017spatial} attempted to reduce this training overhead relying on compressive sensing tools leveraging the note that mmWave channel estimation can be formulated as a sparse reconstruction problem \cite{Alkhateeb2014}. However, compressive sensing based solutions do not take advantage of the available history of the previous channel covariance observations. Further, these techniques still require the stacking of many channel samples to efficiently estimate the channel covariance, which makes them not capable of supporting highly-mobile mmWave applications. In the next section, we present our proposed solution that leverages deep learning tools to address this mmWave channel covariance estimation problem.

\section{Proposed GAN Based Approach} \label{sec:DL_Framework}

In this section, we present our machine learning based mmWave channel covariance prediction algorithm. First, we explain the main idea in \sref{sec:Idea}, before delving into a detailed description of the developed solution and the machine learning modeling in Sections \ref{subsec:Sol} and \ref{subsec:Net}. 

\subsection{The Main Idea}

The key challenge of estimating the channel covariance in highly-mobile mmWave applications is the large training overhead in time, due to the large number of antennas at the transmitters and/or the receivers. Prior research directions in mmWave channel (and channel covariance) estimation used to repeat the estimation process every time they channel (or channel covariance) changes, and did not make use of the previous observations of this estimation process. Intuitively, though, the channel and channel covariance matrices are some functions of the various elements of the environment including the transmitter/receiver locations and scatterers positions. The challenge is that these functions are difficult to characterize analytically as they normally involve many physical interactions and are unique to every environmental setup. Therefore, we propose to leverage the powerful capability of deep learning models to learn this mapping function and enable predicting the mmWave channel covariance matrix given a few features about the channel that are easy to estimate with low training overhead. 

\textbf{Omni-received signals:}  
In \cite{Alkhateeb2018}, the authors showed that when the uplink training pilots are received simultaneously by multiple distributed basestations using omni or quasi-omni antenna patterns, these omni-received signals draw a rich multipath signature for the user location and its interaction with the surrounding environment. This is very interesting as no beam training is needed to acquire these omni-received signals, which dramatically reduces the training overhead. Inspired by this observation, we will adopt the model in which the uplink training pilots are received via only omni patterns, and train the machine learning model to learn the mapping between these omni-received signals and the channel covariance matrix. Mathematically, the omni-received signals are captured by the vector $\by$ in  \eqref{eq:rec_signals} when the combining vectors are set to $\bw_n=[1, 0, ..., 0], \forall n$, i.e., by activating only one receive antenna element at every BS. 

\textbf{Factorized channel covariance:} \label{sec:Idea}
The mmWave channel covariance matrix has normally large dimensions, when employing large number of antennas at the transmitters/receivers. Further, each entry in this covariance matrix can generally take any complex value. This makes it hard for the machine learning models to efficiently predict the channel covariance matrix. Leveraging the sparse nature of the mmWave channels (the existence of only a few dominant paths in the channel), we propose to factorize the channel covariance matrix using the virtual channel model \cite{HeathJr2016}. Noting that $L$ in \eqref{eqn:channel_time} is a finite (and typically small) number, the channel model in \eqref{eqn:channel_time} can be written as
\begin{align}\label{eqn:channel_spatial}
\mathsf{\boldsymbol{h}}_{d,n}=\bU_\mathrm{BS} \mathsf{\boldsymbol{g}}_{d,n},
\end{align}
where $\bU_\mathrm{BS}$ is an $M \times M$ unitary DFT matrix, with the $m$th column equals to $\left[1, e^{- j \frac{2 \pi m}{M}}, e^{- j 2 \frac{2 \pi m}{M}}, ..., e^{-j (M-1) \frac{2 \pi m}{M}} \right]^T$. 
Note that $\mathsf{\boldsymbol{g}}_{d,n}$ is a sparse vector with approximately $L$ nonzero elements corresponding to the $L$ paths of the channel.
Similarly, the frequency domain channel vector at subcarrier $k$ can be written as
\begin{align}
{\bh}_{k,n}= \bU_\mathrm{BS} \bg_{k,n},
\end{align}
with ${\bg}_{k,n}=\sum_{d=0}^{D-1}\mathsf{\boldsymbol{g}}_{d,n}e^{-j\frac{2\pi k}{K}d}$. Finally, the spatial channel covariance, $\bR_{\bh_n}$, can be factorized as
\begin{equation}
\bR_{\bh_n}= \bU_\mathrm{BS} \bR_{\bg_n} \bU_\mathrm{BS}^H, 
\end{equation}
where $\bR_{\bg_n}=\bbE[\bg_{k,n} {\bg}_{k,n}^H]$. In this paper, we will refer to $\bR_{\bg_n}$ as the \textit{virtual} channel covariance. Note that once this virtual channel covariance is estimated, the channel covariance can be directly constructed following $\bR_{\bg_n}= \bU_\mathrm{BS}^H \bR_{\bh_n} \bU_\mathrm{BS}$.  

It is important to note here that the virtual channel covariance matrix, $\bR_{\bg_n}$, is normally a sparse matrix with a few non-zero entries. Therefore, it is much easier to learn the mapping from the omni-received uplink signature $\by$ to the virtual channel covariance matrices as compared to the original channel covariance matrix, $\bR_{\bh_n}$. Consequently, the objective of the machine learning model will be to predict the virtual channel covariance matrix given the omni-received signal. 

\subsection{GAN-Based Channel Covariance Prediction} \label{subsec:Sol}
Our method is inspired by the high resolution image reconstruction which is well performed by GAN. In our task, we interpret the virtual channel covariance matrix, $\bR_{\bg_n}$, as a 2D image. To make sure that the  virtual channel covariance matrix, generated by the GAN network, is directly related to the low-dimensional omni-received signature, we  adopt a conditional GAN network architecture \cite{mirza2014conditional,reed2016generative}. More specifically, in the learning phase, a generative model $G(.)$ takes the omni-received pilot signals $\by$ and a random vector $\bz$ as inputs, and generates an estimated virtual channel covariance matrix $\hat{\bR}_{\bg_n}=G(\bz,\by)$. The discriminative model $D(.)$ then estimates the probability that its input virtual covariance matrix (generated by the generative model) is a real one, given the dataset. The overall loss function condition on $\by$ is defined as
\begin{align}
L(G,D)&=\bbE_{\bR_{\bg_n}}[\log D(\bR_{\bg_n},\by)]\nonumber\\
&+\bbE_{\bz}[\log(1-D(G(\bz,\by)))].
\end{align}
After the model is trained, we use the generator $G$ to directly predict the virtual channel covariance matrix $\hat{\bR}_{\bg_n}$ given the uplink omni-received pilots $\by$.

Note that adopting this GAN network structure is motivated by the dimensions of the output virtual channel covariance matrix, which are much higher than the input training sequences. This is a scenario where traditional multi-layer perceptron networks may not work well \cite{Goodfellow-et-al-2016}. In the following subsection, we describe the adopted conditional GAN network architecture in more detail. 

\subsection{Network Architecture} \label{subsec:Net}

\begin{figure}[t]
	\centerline{
		\includegraphics[width=1\columnwidth]{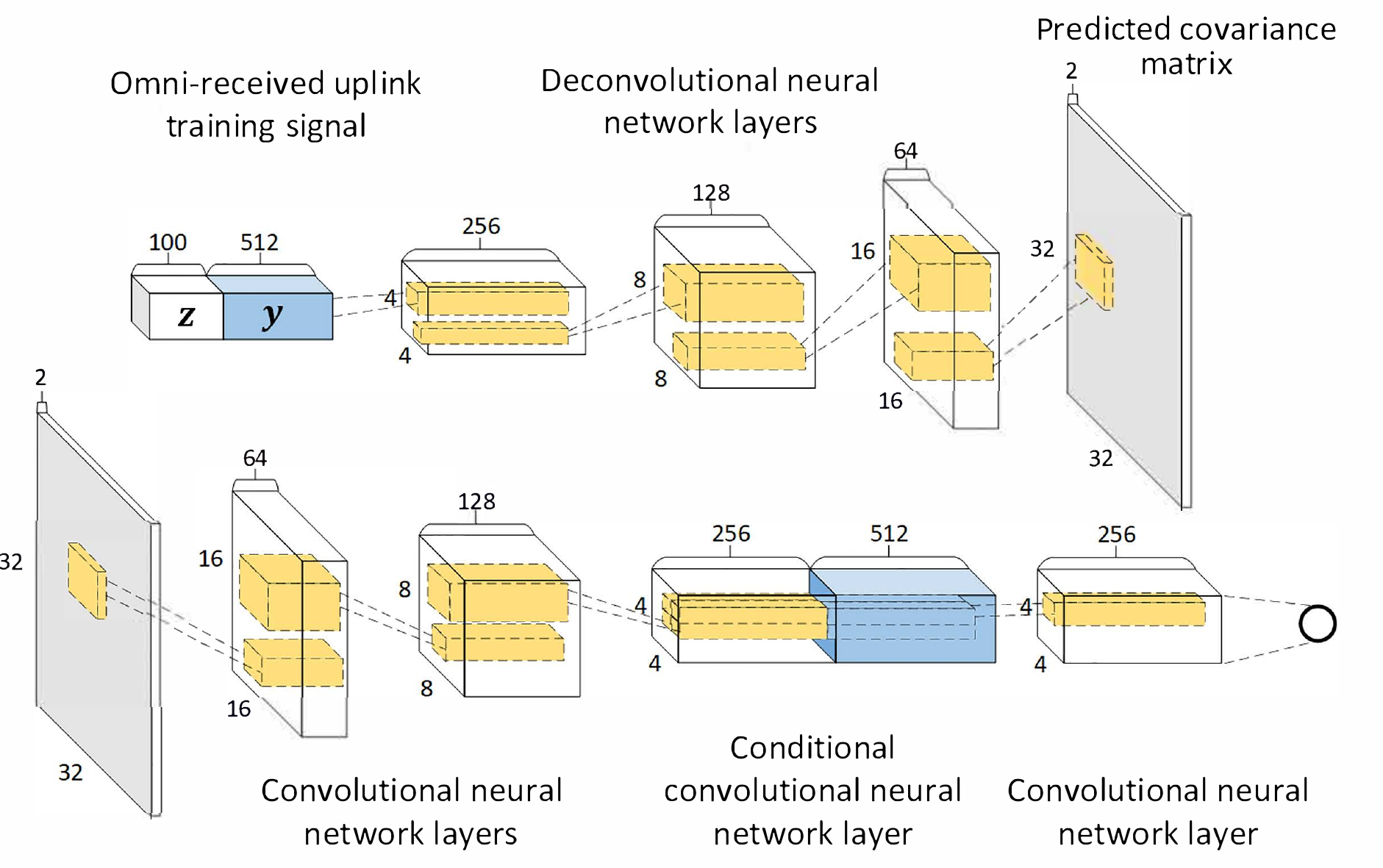}
	}
	\caption{The developed conditional convolutional GAN architecture, with the omni-received training sequences as the input and a predicted covariance matrix as the output. The training sequences, $\by$, are concatenated with a random input vector, $\bz$, in the generator, and with the feature maps in the discriminator. The main function of the discriminator network is to make sure that the generated covariance matrix is tied to the input training sequence.}
	\label{fig:CGAN}
\end{figure}

The considered network architecture which is composed of a generator and discriminator networks is depicted in \figref{fig:CGAN}. The generator network is denoted as $G:\mathbb{R}^Z\times\mathbb{R}^{NK}\rightarrow\mathbb{R}^{M^2}$ while the discriminator is represented by $D:\mathbb{R}^{M^2}\times\mathbb{R}^{NK}\rightarrow \{0,1\}$. Here, $Z$ denotes the dimension of the noise input to $G$, and  $NK$ and $M^2$ are the dimensions of the training sequences and the covariance matrix. We input both the real and imaginary parts of the omni-received sequence $\by$. In the example of \figref{fig:CGAN}, we consider a setup with 4 coordinating BSs and with $K=64$ subcarriers. Therefore, the size of the input, accounting for both the real and imaginary parts, equals $512$, in addition to a noise vector $\bz$ of size 100. 

In the generator $G$, we first generate a noise vector $\bz$ whose elements are drawn from $\cN(0,1)$, then concatenate the training sequences $\by$ to $\bz$. Following this, the estimation process is a deconvolutional network. By feed-forwarding the concatenated input, $G$ generates an estimated virtual channel covariance matrix $\hat{\bR}_{\bg_n}$ via $\hat{\bR}_{\bg_n}\leftarrow G(\bz,\by)$. Therefore, the estimated virtual covariance matrix is generated through the estimation process of $G$ conditioned on the training sequences and the random input vector. Note that the example in \figref{fig:CGAN} assumes that each BS is equipped with 32-element ULA. Thus, the covariance matrices have dimensions $32 \times 32$. 

In the discriminator $D$, several layers of stride-2 convolution with spatial batch normalization followed by leaky ReLU are applied. Then, a  conditional CNN layer (with spatial dimensions of $4 \times 4$ in the example of \figref{fig:CGAN}) is added and the training sequences are concatenated in the depth dimension. Finally, a convolution followed by a rectification is performed to make the final output conditioned on the training sequence. Note that the conditional CNN layers combine the predicted virtual channel covariance matrix and its corresponding training sequences in a straightforward way by concatenating them and filtering the concatenated results. Such a structure can merge the information carried by both the covariance matrix and the training sequence \cite{Reed2016,reed2016generative}, which enhances the dependence of the discriminator's output on the training sequences. 

\section{Simulation Results} \label{sec:Results}
In this section, we describe in detail the simulation setup including the channel models, datasets generation, and GAN model parameters, as well as the simulation results.

\begin{figure}[t]
	\centerline{
		\includegraphics[width=8.5cm]{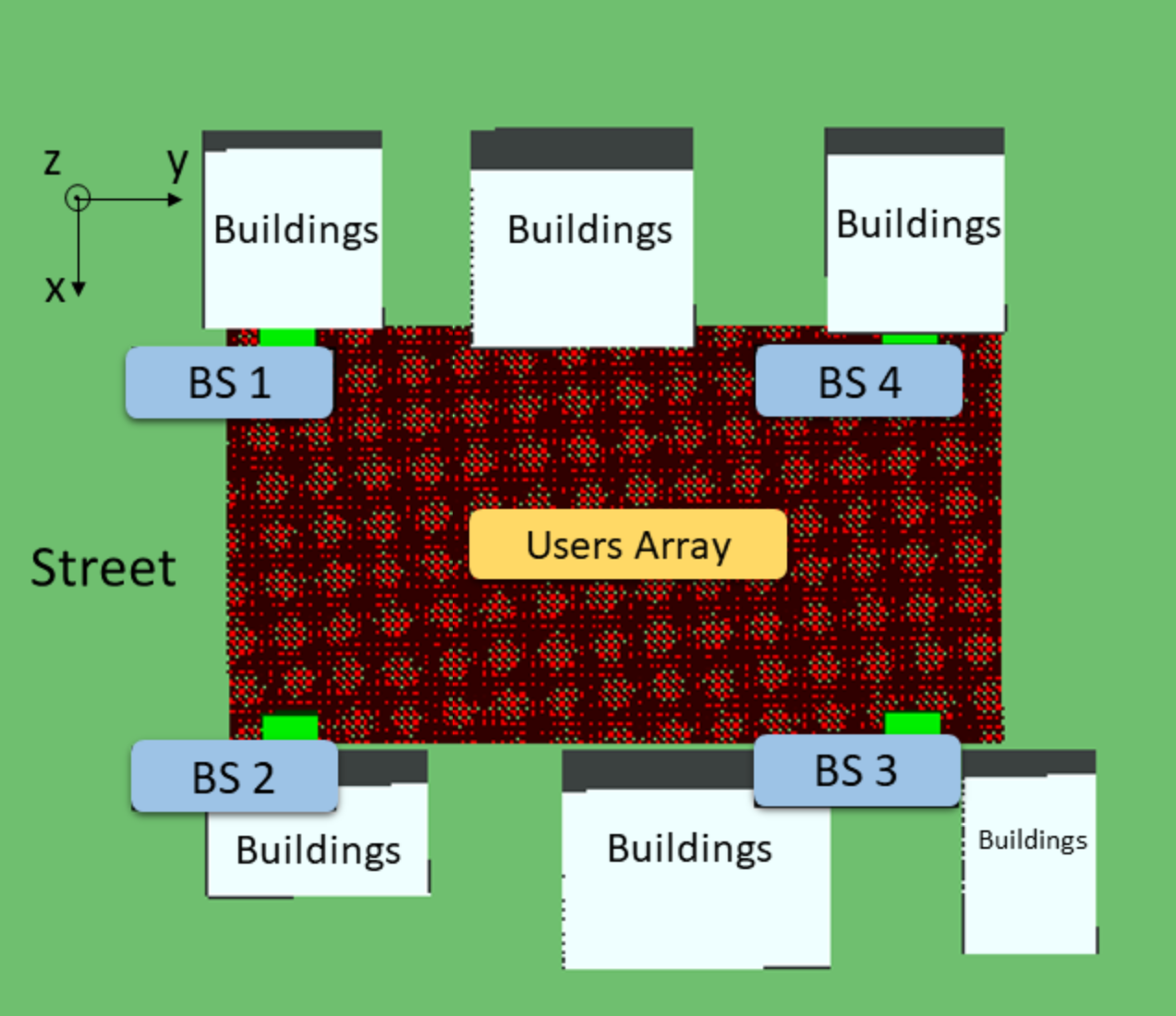}
	}
	\caption{The street model where $N=4$ BSs are installed on 4 lamp posts on both sides of a street and receive uplink training from one user simultaneously. The 4 lamps are located on the corners of a 30m (x-axis, along the street) by 20m (y-axis, across the street) rectangle. Each BS has a ULA with $M$ antennas along the y-axis and is installed at height 6m while the vehicular user has a single antenna at height 1m.}
	\label{fig:Street}
\end{figure}

\textbf{System and channel models:} We adopt the mmWave system and channel models in \sref{sec:Model}, with the channel parameters (angles of arrival, path loss, etc.) generated using the commercial ray-tracing simulator Wireless InSite \cite{Remcom}. The system considers 4 BSs serving one mobile over the 60GHz band. The 4 BSs are installed on 4 lamp posts at both sides of a street, as shown in Fig. \ref{fig:Street}. The 4 lamps are located on the corners of a 30m (x-axis, along the street) by 20m (y-axis, across the street) rectangle. Each BS has a ULA with $M=32$ antennas along the y-axis and is installed at height 6m while. For simplicity, the vehicular user is equipped with a single antenna at height 1m. 

To simulate the channels between the BSs and the user at different locations of the street, we consider an x-y rectangular grid of candidate antenna/user locations with $200 \times 300 = 60k$ points. For every candidate antenna location, an uplink training signal is transmitted by the mobile user and is received simultaneously by the 4 BS using both omni-pattern (one active antenna) and the ULA. The omni-received signals are concatenated to form the omni-received uplink training sequence, $\by$, which is used as an input to the machine learning model. The signals received by the ULA are used to construct the virtual channel covariance matrices, $\bR_{\bg_n}$, which are the outputs of the machine learning model. Note that every candidate user location results in one point in the machine learning dataset, which consists of a training sequence and a virtual channel covariance matrix. The number of the candidate antennas/users then decides the size of the database. The different channel parameters (AoAs, path gains, etc.) are generated by the ray-tracing simulator and are used to construct the virtual channel covariance matrices in MATLAB. We then train the GAN network at BS 1 which is placed at the top left corner of the rectangular street area to predict the channel covariance matrix at this BS given the training sequence, $\by$, collected from the 4 coordinating BSs. 

\textbf{Machine learning model:} In the GAN network architecture, we set $\bz=100$ and $K=64$. Therefore, the dimensions of the input, which concatenates the random vector $\bz$ and the real/imaginary training sequence, equals $1 \times 612$. We treat the virtual channel covariance matrix as a gray image with dimensions $M \times M \times2$. Further, a normalization is performed such that both the omni-received sequences (the inputs) and the virtual covariance matrices (the outputs) are normalized by the maximum absolute value of their elements. Alternating steps of updating the generator and the discriminator network are used. The learning rate is set to 0.0002. We use the ADAM optimizer with momentum 0.5 and a batch size of 256. Finally, the network is trained for 200 epochs. Our machine learning model implementation was built by TensorFlow.

\begin{figure}[t]
	\centerline{
		\includegraphics[width=1.1\columnwidth, trim={0pt, 0pt, 0pt, 50pt}]{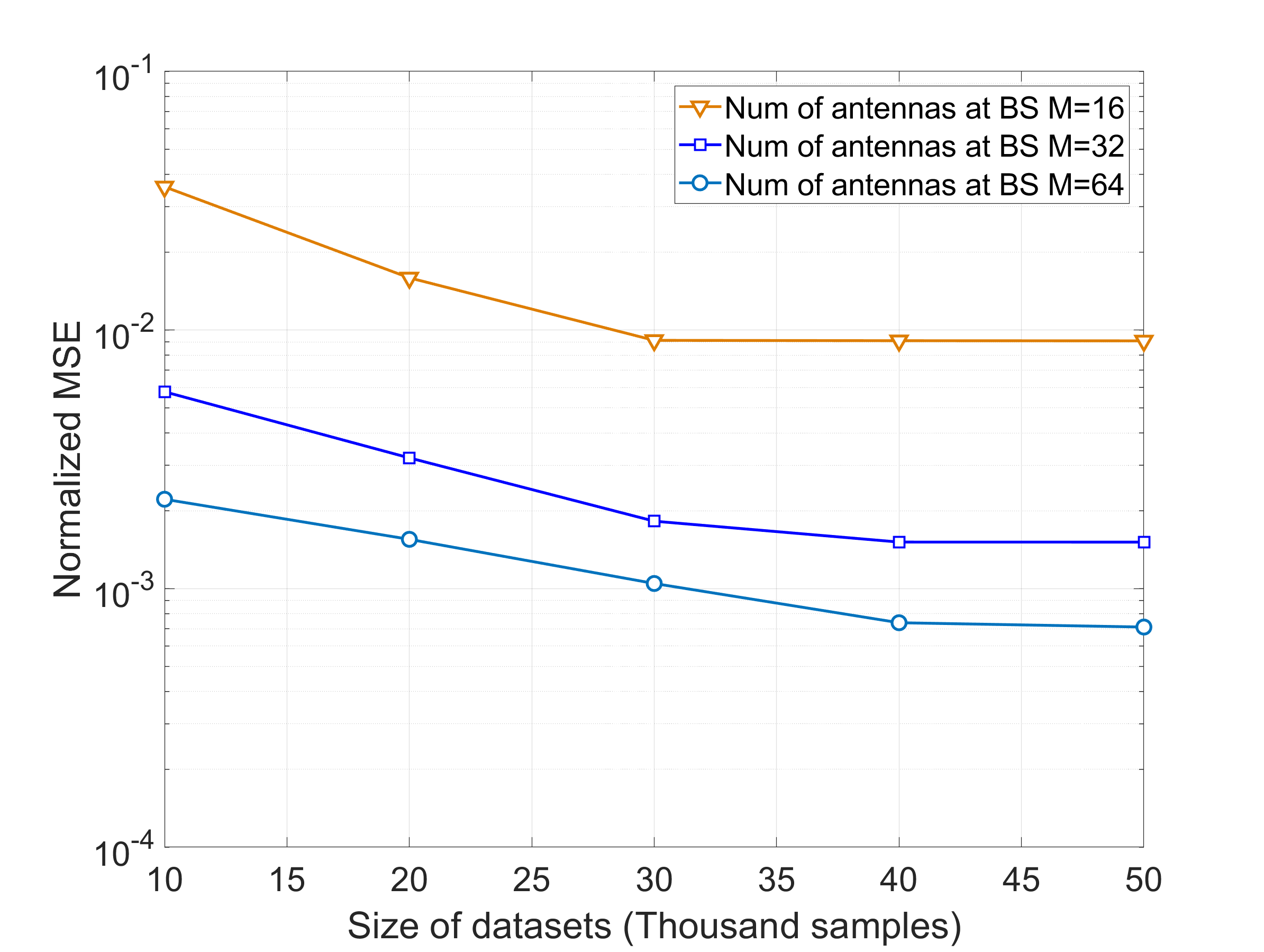}
	}
	\caption{The NMSE performance of the estimated covariance matrices using the proposed GAN-based solutions. The performance with large array sizes is generally better than that with smaller arrays. This is intuitive because we treat the channel covariance matrices as images and use convolutional neural networks to exploit their features. These networks are normally able to exploit more features with larger images.}
	\label{fig:mse}
\end{figure}

\textbf{Performance evaluation:} To evaluate the performance of the proposed solution for predicting the mmWave virtual channel covariance matrices, we use the normalized mean square error (NMSE), defined as
\begin{align}
\rm{NMSE}=\frac{||\hat{\bR}_{\bg_n}-\bR_{\bg_n}||_F^2}{||\bR_{\bg_n}||_F^2}.
\end{align}

In Fig. \ref{fig:mse}, the average NMSE of the predicted virtual channel covariance matrices is plotted versus the size of the training dataset for different sizes of the BS antenna array. This figure shows that the performance of the proposed solution is better with large array sizes. This is intuitive because we treat the channel covariance matrices as images and use CNNs in the discriminator to exploit their multipath features. CNNs are normally able to exploit more features for larger images, which is the case with large antenna arrays in our problem. 

\begin{figure}[t]
	\centerline{
		\includegraphics[width=1.05\columnwidth]{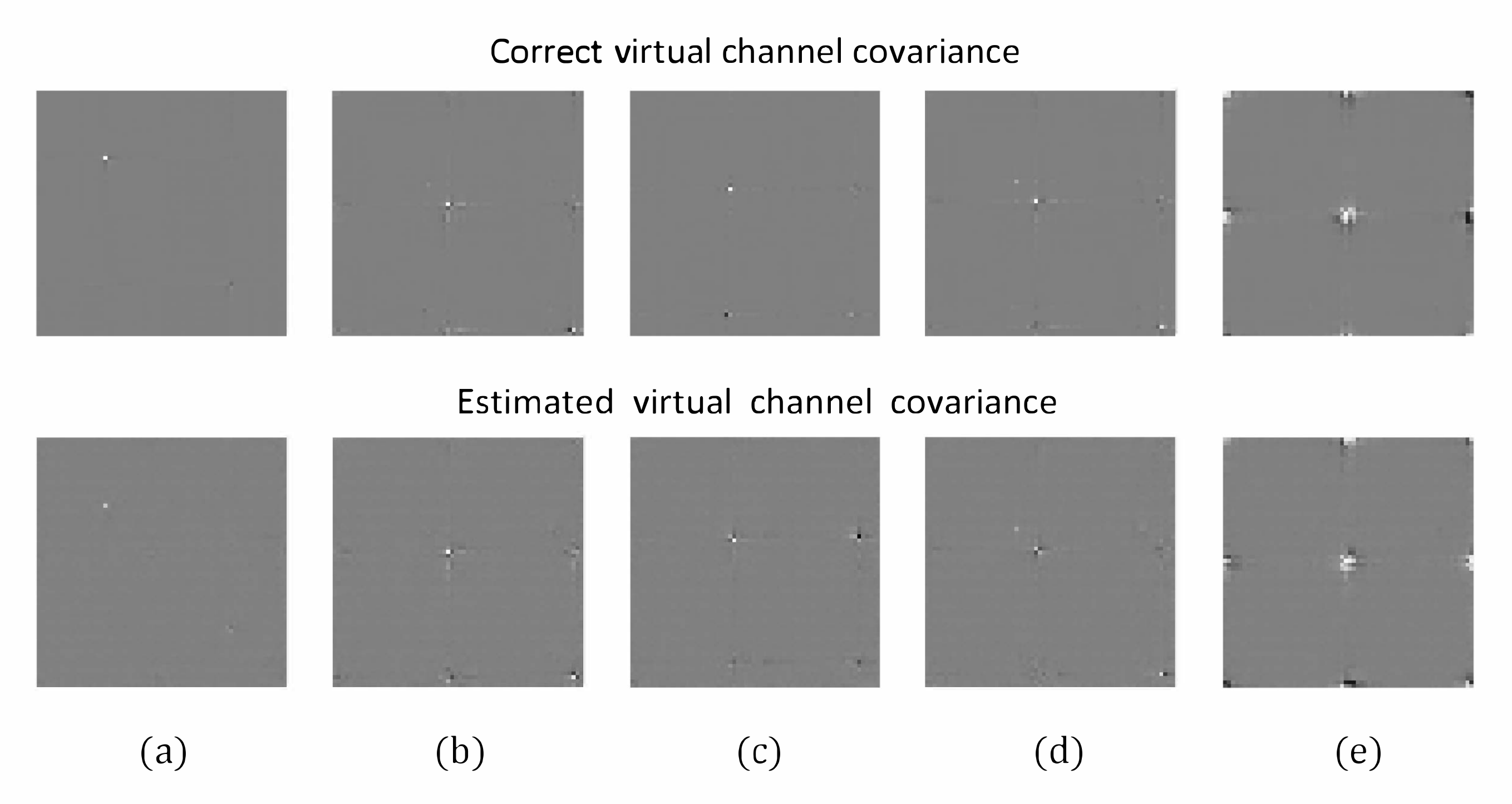}
	}
	\caption{Example results of the real parts of estimated channel covariance shown as images at BS1. Our GAN model successfully estimates the channel covariance matrices for the different cases: when only one strong path exists (as in (a)), when two path exist (as in (b) and (c)), and when more than two paths exist (as in (d) and (e)).}
	\label{fig:examples}
\end{figure}

Since we treat the virtual channel covariance matrices as images, it is interesting to visually compare the original and estimated virtual covariance matrices. In \figref{fig:examples}, we plot the results of the real parts of the original and estimated covariance matrices at BS 1. The white points in the main diagonal of the images represent strong receiving paths. The brightness of the points in the off-diagonal positions illustrates the level of correlation among the different paths. For example, the brightness of entry $(i,j)$ reflects the strength of correlation of the $i$th and $j$th paths. 

Note that we only considered the 5 strongest paths for every BS-user channel when generating the dataset via the ray-tracing simulator. Therefore, the covariance matrices have a small number of paths (bright points).  \figref{fig:examples} shows that our GAN model can successfully estimate the virtual channel covariance matrices for different cases: when only one strong path exists (Fig. \ref{fig:examples} (a)), two path exist (Fig. \ref{fig:examples} (b), (c)), and more than two paths exist (Fig. \ref{fig:examples} (d), (e)).

\section{Conclusion} \label{sec:Conclusion}
In this paper, we developed a novel mmWave channel covariance estimation/prediction solution based on recent deep learning techniques. The proposed solution learns the mapping between the uplink signals received simultaneously at multiple BSs using only omni-patterns and the covariance matrices. This solution, therefore, requires negligible time overhead in estimating the channel covariance matrices. In our machine learning model, we treat the covariance matrices as images and leverage conditional generative adversarial networks to learn the important features of these images. Simulations results, based on accurate ray-tracing and practical deployment scenarios, showed that the developed deep learning based solution efficiently predicts the mmWave channel covariance matrices with small mean-squared errors. In the future work, it is interesting to extend the current results to multi-user scenarios and to the cases where both the BSs and mobile users are equipped with antenna arrays. 




\begin{thebibliography}{10}
	\providecommand{\url}[1]{#1}
	\csname url@samestyle\endcsname
	\providecommand{\newblock}{\relax}
	\providecommand{\bibinfo}[2]{#2}
	\providecommand{\BIBentrySTDinterwordspacing}{\spaceskip=0pt\relax}
	\providecommand{\BIBentryALTinterwordstretchfactor}{4}
	\providecommand{\BIBentryALTinterwordspacing}{\spaceskip=\fontdimen2\font plus
		\BIBentryALTinterwordstretchfactor\fontdimen3\font minus
		\fontdimen4\font\relax}
	\providecommand{\BIBforeignlanguage}[2]{{%
			\expandafter\ifx\csname l@#1\endcsname\relax
			\typeout{** WARNING: IEEEtran.bst: No hyphenation pattern has been}%
			\typeout{** loaded for the language `#1'. Using the pattern for}%
			\typeout{** the default language instead.}%
			\else
			\language=\csname l@#1\endcsname
			\fi
			#2}}
	\providecommand{\BIBdecl}{\relax}
	\BIBdecl
	
	\bibitem{Alkhateeb2018}
	A.~Alkhateeb, S.~Alex, P.~Varkey, Y.~Li, Q.~Qu, and D.~Tujkovic, ``Deep
	learning coordinated beamforming for highly-mobile millimeter wave systems,''
	\emph{IEEE Access}, vol.~6, pp. 37\,328--37\,348, 2018.
	
	\bibitem{Adhikary2014}
	A.~Adhikary, E.~A. Safadi, M.~K. Samimi, R.~Wang, G.~Caire, T.~S. Rappaport,
	and A.~F. Molisch, ``Joint spatial division and multiplexing for mm-wave
	channels,'' \emph{IEEE Journal on Selected Areas in Communications}, vol.~32,
	no.~6, pp. 1239--1255, June 2014.
	
	\bibitem{park2017spatial}
	S.~Park and R.~W. Heath~Jr, ``Spatial channel covariance estimation for the
	hybrid {MIMO} architecture: A compressive sensing based approach,''
	\emph{arXiv preprint arXiv:1711.04207}, 2017.
	
	\bibitem{lee2016channel}
	J.~Lee, G.-T. Gil, and Y.~H. Lee, ``Channel estimation via orthogonal matching
	pursuit for hybrid {MIMO} systems in millimeter wave communications,''
	\emph{IEEE Transactions on Communications}, vol.~64, no.~6, pp. 2370--2386,
	2016.
	
	\bibitem{HeathJr2016}
	R.~W. Heath, N.~González-Prelcic, S.~Rangan, W.~Roh, and A.~M. Sayeed, ``An
	overview of signal processing techniques for millimeter wave {MIMO}
	systems,'' \emph{IEEE Journal of Selected Topics in Signal Processing},
	vol.~10, no.~3, pp. 436--453, April 2016.
	
	\bibitem{alkhateeb2016frequency}
	A.~Alkhateeb and R.~W. Heath, ``Frequency selective hybrid precoding for
	limited feedback millimeter wave systems,'' \emph{IEEE Transactions on
		Communications}, vol.~64, no.~5, pp. 1801--1818, 2016.
	
	\bibitem{Alkhateeb2014}
	A.~Alkhateeb, O.~El~Ayach, G.~Leus, and R.~Heath, ``Channel estimation and
	hybrid precoding for millimeter wave cellular systems,'' \emph{IEEE Journal
		of Selected Topics in Signal Processing}, vol.~8, no.~5, pp. 831--846, Oct.
	2014.
	
	\bibitem{mirza2014conditional}
	M.~Mirza and S.~Osindero, ``Conditional generative adversarial nets,''
	\emph{arXiv preprint arXiv:1411.1784}, 2014.
	
	\bibitem{reed2016generative}
	S.~Reed, Z.~Akata, X.~Yan, L.~Logeswaran, B.~Schiele, and H.~Lee, ``Generative
	adversarial text to image synthesis,'' \emph{arXiv preprint
		arXiv:1605.05396}, 2016.
	
	\bibitem{Goodfellow-et-al-2016}
	I.~Goodfellow, Y.~Bengio, and A.~Courville, \emph{Deep Learning}.\hskip 1em
	plus 0.5em minus 0.4em\relax MIT Press,
	\url{http://www.deeplearningbook.org}.
	
	\bibitem{Reed2016}
	S.~E. Reed, Z.~Akata, S.~Mohan, S.~Tenka, B.~Schiele, and H.~Lee, ``Learning
	what and where to draw,'' in \emph{Advances in Neural Information Processing
		Systems}, 2016, pp. 217--225.
	
	\bibitem{Remcom}
	Remcom, ``Wireless insite,'' \url{http://www.remcom.com/wireless-insite}.
	
\end{thebibliography}
\end{document}